\documentclass[11pt,a4paper]{article}
\usepackage{jheppub}
\usepackage{mathtools}
\usepackage{enumitem}
\usepackage{bm}
\usepackage{physics}
\usepackage{sansmath}
\usepackage{hyperref}
\usepackage{cleveref}

\SetMathAlphabet{\mathsfsl}{sans}{\sansmathencoding}{\sfdefault}{m}{n}

\arxivnumber{2403.10611}

\title{\boldmath Two-dimensional gauge anomalies and $p$-adic numbers}
\author[a]{Imogen Camp,}
\author[b]{Ben Gripaios}
\author[c]{and Khoi {Le Nguyen Nguyen}}

\affiliation[a]{Rudolf Peierls Centre for Theoretical Physics, University of Oxford, Parks Road, Oxford, OX1 3PU, United Kingdom}

\affiliation[b]{Cavendish Laboratory, University of Cambridge, J.J. Thomson Avenue, Cambridge, CB3 0HE, United Kingdom}

\affiliation[c]{DAMTP, University of Cambridge, Wilberforce Road, Cambridge, CB3 0WA, United Kingdom}

\emailAdd{imogen.camp@some.ox.ac.uk}
\emailAdd{gripaios@hep.phy.cam.ac.uk}
\emailAdd{kl518@cam.ac.uk}

\abstract{We show how methods of number theory can be used to study anomalies in gauge quantum field theories in spacetime dimension two. To wit, the anomaly cancellation conditions for the abelian part of the local anomaly admit solutions if and only if they admit solutions in the reals and in the $p$-adics for every prime $p$ and we use this to build an algorithm to find all solutions.}

\begin{document}
	\maketitle
	\flushbottom

	\section{Introduction}\label{section:intro}
        Two-dimensional gauge theories, already important in their own right for condensed matter physics ({\em cf.  e.g.}\cite{Altland_Simons_2010a}), often provide a sweet spot for high-energy physics, as the rich phenomena of
        four-dimensional gauge theories remain present but become much more tractable.
		Examples include dynamical mass generation \cite{Schwinger:1962tp}, confinement \cite{tHooft:1974pnl,Callan:1975ps}, chiral symmetry breaking \cite{Zhitnitsky:1985um,PhysRevD.37.946}, non-perturbative studies of chiral gauge theories using a spacetime lattice \cite{EICHTEN1986179,PhysRevLett.116.211602,GOLTERMAN2001189,seifnashri2024liebschultzmattis,PhysRevLett.128.185301} and gapped chiral fermions \cite{PhysRevB.107.014311,PhysRevD.99.111501,Tong_2022}. Here we describe yet another example of this, by studying the phenomenon of gauge theories that may not be theories at all, because they suffer from anomalies.

        Before discussing why studying anomalies in two-dimensional theories is easy (or easier), let us discuss the sense in which studying them in four dimensions is hard. To do so, it is helpful to frame the discussion in terms of questions that a physicist might wish to answer. An obvious first question is: given a gauge group and fermion representation (say, the Standard Model (SM) with its three generations of quarks and leptons), is it anomalous or not? As the example of the SM shows \cite{Garcia-Etxebarria:2018ajm,Davighi_2020}, the answer (which is no, thank goodness) requires no small amount of algebro-topological jiggery-pokery.

        This first question becomes vastly easier if one considers only so-called local anomalies, associated to the Lie algebra of the gauge group, since it reduces to checking that certain polynomials evaluate to zero \cite{BOUCHIAT1972519}. But even local anomalies prove challenging if we ask further questions relevant to the search for new physics, such as: starting from a given gauge group and fermion representation, in what ways can we extend the group and/or the representation? Such questions are important in, {\em e.g.}, attempting to unify the SM gauge couplings or explain the pattern of fermion masses and mixings using flavour symmetries.

        Here again, the problem can be broken down to a certain extent, using the decomposition of the Lie algebra into its semisimple and abelian summands. Ref.~\cite{Allanach:2021bfe}, for example, solves the problem of finding all of the semisimple algebras that achieve unification, in the sense that they contain the SM algebra. 

        The remaining part, which we might call nonsemisimple since it involves both the abelian and semisimple summands, is much harder. Indeed, it amounts to a famously hard problem of number theory, namely finding integral (or equivalently rational) solutions of (multiple) polynomial equations up to cubic order.
But here too there have been notable recent successes, such as the general solution for the possible charges of any number of fermions in a theory with gauge group $U(1)$ \cite{PhysRevLett.123.151601,Allanach2019GeometricGS} and the general solution for the possible charges that can be carried by the fermions of the Standard Model (along with 3 right-handed neutrinos added to give neutrinos their observed masses) under a single extra $U(1)$ gauge factor \cite{PhysRevLett.125.161601}. These methods, relying on {\em ad hoc} variations on the method of chords, a number-theoretic tool going back at least to Fermat, are of limited validity. If we swap $U(1)$ for $U(1) \times U(1)$ in the first example, or decrease the number of right-handed neutrinos from three to two (which is consistent with data, given that we only observe mass differences), then no method of solution is known. 

In this work, we study this `hardest part of the problem' for two-dimensional gauge theories. A first win is that because there are no mixed gauge anomalies, this nonsemisimple part essentially collapses to an abelian part \footnote{More precisely, the representations of the semisimple summand of the Lie algebra enter only via their degrees.}. Here too, the method of chords turns out to be of limited applicability, but we will see that a different number-theoretic tool, namely the theory of quadratic forms on rational vector spaces, allows us to give a complete solution to any question we might wish to ask. The key difference is that the polynomial equations to be solved in two dimensions are all quadratic, rather than cubic or higher; for such equations the local-to-global principle of Hasse and Minkowski (to whom we owe much more than just a metric!) applies, meaning that they admit solutions in the rationals if and only if they admit solutions in every completion of the rationals, {\em i.e.} in the reals and in the $p$-adic numbers for every prime $p$. The latter conditions are relatively easy to determine, by finding invariants that classify the possible quadratic forms.
 One such invariant is the Witt index, which has a direct physical interpretation as the solution to another problem whose solution for the SM remains unknown: what is the maximum number of $U(1)$ gauge factors that can be added?  
For any given number of $U(1)$ factors up to the maximum, we describe an algorithm
to find all possible values of the corresponding charges. Happily, this requires no explicit knowledge of $p$-adics. An ancillary file implements it as a {\tt Mathematica} notebook.
	\section{Sketch of the problem \& solution}\label{section:sketch}
        Before getting into the nitty-gritty, let us discuss some simple examples of the kinds of equation we wish to solve. Any fool can see that $x_1^2 + x_2^2 = 0$ and $x_1^2 - 2x_2^2 = 0$ have no non-trivial rational solutions, but it takes a Hasse or a Minkowski to observe that this is implied by the fact that they have no solutions in the reals or the 2-adics, respectively. These examples are uninteresting for physics since, as we shall see, a possible gravitational anomaly forces the coefficients in such equations to sum to zero. So consider instead the equation
$$2x_1^2+3x_2^2-4x_3^2-x_4^2=0.$$
As we shall later see, it is easy to cook up a gauge theory with a single $U(1)$ factor in which each of $x_1,x_2,x_3$ and $x_4$ corresponds to the $U(1)$ charge of a chiral fermion (either left-moving or right-moving, according to the sign of the coefficient) carrying some irreducible representation (henceforth, `irrep') of the gauge group, where this is the only equation to be satisfied to guarantee that the theory is free of local anomalies. We obviously have solutions in which $x_1=x_2=x_3=x_4$ is a rational multiple of unity, from which it is straightforward to find all solutions ({\em e.g.} by the method of chords), each of which lies on a one-dimensional vector subspace of solutions in the vector space $\mathbb{Q}^4 \ni (x_1,x_2,x_3,x_4)$. But suppose we consider instead the gauge theory with two $U(1)$ factors. Now an anomaly-free theory is specified by a two-dimensional vector subspace of solutions in $\mathbb{Q}^4$. Such vector subspaces certainly exist over the reals, as can easily be seen by doing the change of variables $(x_1,x_2,x_3,x_4) \mapsto (x_1/\sqrt{2},x_2/\sqrt{3},x_3/2,x_4)$, but this is illegal over the rationals. So does it have any solutions over the rationals? It turns out that the answer is no, 
as can be seen by studying the solutions in, {\em e.g.}, the $2$-adics. 

This example makes clear how to proceed in the analysis of local anomalies in a general gauge theory. In each case we must find all $m$-dimensional vector subspaces of the rational vector space $\mathbb{Q}^n \ni x_i$, which are `totally isotropic' (to use the lingo that we introduce later), in that each vector in them satisfies the equation
\begin{equation}
	\sum_{i=1}^n c_i x_i^2 = 0.\label{equation:quaddiophantine}
\end{equation}
In physics terms, $m$ is the number of abelian gauge fields, $n$ is the number of irreps of the fermions, the $c_i$ are non-zero integers related to the degrees and multiplicities of the irreps, and the equation corresponds to cancellation of the local abelian anomaly.
The $c_i$ must further satisfy $\sum_{i=1}^n c_i = 0$, corresponding to cancellation of the local gravitational anomaly. There are no anomalies mixing gravitation and gauge fields, nor anomalies mixing the abelian gauge fields and the semisimple ones. (Thus the only possible remaining local anomaly is associated purely to the semisimple summand and will be discussed elsewhere \cite{gripaios_toappear}.)

The first step in the analysis is to determine the Witt index $w$ (an algorithm is given in \cite{Beale1989}),
which corresponds to the maximum possible value of $m$ and hence the maximal number of abelian gauge fields. Witt's theory of quadratic forms tells us that every totally isotropic subspace is contained in a maximal one,
so by finding the maximal ones, we can find all solutions with any number of abelian gauge fields.

The second step is to observe that we can find all maximal ones if we can find just one of them, since any two are connected by a generalized notion of orthogonal transformation. Such transformations can be simply parameterized using a notion of a generalized skew-symmetric matrix, via a Cayley transformation. 

It remains for us to find a single maximal totally isotropic subspace. Here too, Witt's theory comes to the rescue: it shows that such a space can be constructed iteratively in $w$ steps by finding at each stage a single vector $y_j \in \mathbb{Q}^{n^\prime}$ such that $\sum_{j=1}^{n^\prime} b_j y_j^2 = 0$, where the value of $n^\prime$ goes down by 2 at each step, and the coefficients $b_j$ change too (in particular, they no longer sum to zero, necessarily). Such a vector (which exists!) can be found by simple trial-and-error, though more sophisticated algorithms, such as {\tt Mathematica}'s {\tt FindInstance}, are available.
	\section{Details}\label{section:nit}
        {\em The local anomaly and its abelian part.} Consider a quantum field theory in two-dimensional Minkowski space whose gauge group is a compact Lie group $G$ with Lie algebra  $\mathfrak{g}$ \footnote{In what follows, formul\ae\ more familiar to physicists can be obtained by choosing an explicit basis for $\mathfrak{g}$.}. 
Given right-moving fermions carrying a finite-dimensional representation of $G$ inducing a representation $R: \mathfrak{g} \to \mathrm{End}\ V$ of $\mathfrak{g}$, 
a Feynman diagram computation with a loop of fermions and a pair of  either gauge boson or graviton external legs shows that there is a purely gauge local anomaly proportional to $\mathrm{tr}\ R^2$ and a purely gravitational local anomaly proportional to $\mathrm{tr}\ 1_V$, where $1_V$ denotes the identity map on $V$. The expressions for left-moving fermions are similar, but opposite in sign, and consistency of the theory requires that the sums of the left- and right-moving contributions vanish.

Now, the component of $G$ that is connected to the identity can be written as $(A \times S)/Z$, where $A$ is an $m$-dimensional torus, $S$ is simply-connected, and $Z$ is a finite group; correspondingly, 
$\mathfrak{g}= \mathfrak{s} \oplus \mathfrak{a}$ is a sum of a semisimple summand $\mathfrak{s}$ and an abelian summand $\mathfrak{a}$ \cite{HofmannKarlHeinrich2013Tsoc}. Decomposing $R$ into irreps of $\mathfrak{g}$, each of which is a tensor product of irreps of $\mathfrak{s}$ and $\mathfrak{a}$, and using the fact that representations of $\mathfrak{s}$ are traceless, it follows that the purely gauge anomaly is the sum of a summand that is constant on $\mathfrak{a}$, which we call the semisimple local anomaly, and another constant on $\mathfrak{s}$ \footnote{This result does not hold in higher spacetime dimensions, where there are also mixed gauge anomalies.}. This latter summand, which we call the abelian local anomaly, may be written explicitly as $\sum_i c_i R_i^2$, where $i \in \{1,\dots,n\}$ indexes the inequivalent irreps of $\mathfrak{g}$ that appear \footnote{Irreps of $G$ do not necessarily descend to irreps of $\mathfrak{g}$, but rather may be reducible. {\em E.g.} the defining irrep of the Lie group $O(2)$ has degree 2 and the Lie group $\Sigma_n \times U(1)$, where $\Sigma_n$ is the group of permutations on $n$ objects has an irrep of degree $n-1$, but these are necessarily reducible representations of the Lie algebra $\mathfrak{u}(1)$.}, the linear map $R_i: \mathfrak{a} \to \mathbb{R}$ denotes the corresponding irrep of $\mathfrak{a}$ (necessarily of degree one), and $c_i \in \mathbb{Z}\setminus\{0 \}$ is the product of the degree of the corresponding irrep of $\mathfrak{s}$, the multiplicity with which the irrep $i$ appears, and a plus (resp. minus) sign if the corresponding fermion is left- (resp. right)-moving. 

It is easy to come up with a gauge theory for which each $c_i$ can be chosen to be arbitrary integers whilst ensuring that the semisimple part of the anomaly vanishes \footnote{Set $G=\protect\prod_i\Sigma_{c_i+1} \times U(1)$ and ask for $n$ left-movers with the $i$th left-mover carrying the degree $c_i$ irrep of $\Sigma_{c_i+1}$ and the trivial irrep of the other factors.}. So in general we need to be able to solve cases in which the $c_i$ sum to zero (to satisfy the gravitational anomaly constraint), but are otherwise arbitrary.

{\em The quadratic space.} Since the $c_i$ are integral, they can be used to define (up to an irrelevant ordering ambiguity) a diagonal quadratic form $q$ on the vector space $\mathbb{Q}^n$ over $\mathbb{Q}$. Since $A$ is an $m$-dimensional torus, the irreps $R_i$ of $\mathfrak{a}$ that can appear define a lattice in the dual space of $\mathfrak{a}$ that is isomorphic (as a group) to $\mathbb{Z}^m$.
Without loss of generality, we may demand that $\mathfrak{a}$ act faithfully on the fermions (if it doesn't, the corresponding gauge bosons will be decoupled from the fermions and we might as well replace $\mathfrak{a}$ by the largest subalgebra that does act faithfully). If the local abelian anomaly is to vanish, it is then necessary and sufficient that the $n$ elements in $\mathbb{Z}^m$ corresponding to each of the $R_i$ define an $m$-dimensional linear subspace $\mathbb{Q}^m \subset \mathbb{Q}^n$ on which $q$ vanishes. Conversely, given any $m$-dimensional linear subspace $\mathbb{Q}^m \subset \mathbb{Q}^n$ on which $q$ vanishes, we can clear denominators to find $R_i$ such that the anomaly vanishes.
We now describe how to find such subspaces.

{\em Maximal isotropic subspaces.} We introduce some standard vocabulary of quadratic spaces \cite{CasselsJ.W.S.JohnWilliamScott1978Rqf,LamT.Y.Tsit-Yuen2005Itqf,SerreJean-Pierre1973Acia}. For a field $\Bbbk$ of characteristic $\neq2$ a {\em quadratic space} $(V,q)$ consists of a vector space $V$ over $\Bbbk$ along with a symmetric bilinear map $q: V \times V \to \Bbbk$, which defines a quadratic form via $\vb{v} \mapsto q(\vb{v}):= q(\vb{v}, \vb{v})$. Choosing an ordered basis $(\vb{e}_k)$ of $V$ defines a {\em Gram matrix} $\mathsf{Q}$ whose $kl$-th entry is $q(\vb{e}_k, \vb{e}_l)$.
We call $(V,q)$ {\em regular} if the determinant of $\mathsf{Q}$ does not vanish (in which case it defines a basis-independent element in $\Bbbk^\times/(\Bbbk^\times)^2$). Since $c_i \neq 0$ in the discussion above, all quadratic spaces considered will be regular.

A non-zero vector $\vb{b}\in V$ is said to be \emph{isotropic} if $q(\vb{b})=0$. 
A regular $(V,q)$ is \emph{isotropic} if it has an isotropic vector and \emph{anisotropic} if it does not. It is \emph{totally isotropic} if every non-zero vector is isotropic. 
A subspace $U \subset V$ is \emph{maximal totally isotropic} if $(U,q_U)$ (where $q_U$ denotes the restriction of $q$ to $U$) is totally isotropic and $V$ has no other totally isotropic subspace containing $U$. 

An important example of an isotropic space is a \emph{hyperbolic plane}, which is a two-dimensional quadratic space $(V,q)$ along with a basis $\vb{v}_1, \vb{v}_2$ for $V$ satisfying ${q(\vb{v}_1)=q(\vb{v}_2)=0}$ and $q(\vb{v}_1,\vb{v}_2)=1$. In fact, every isotropic space $(V,q)$ with some isotropic vector $\vb{v}_1$ contains a hyperbolic plane with basis $\vb{v}_1$ and
\begin{equation}
	\vb{v}_2=\frac{1}{q(\vb{w},\vb{v}_1)}\left[\vb{w}-\frac{q(\vb{w})}{2q(\vb{w},\vb{v}_1)}\vb{v}_1\right],\label{equation:v2}
\end{equation}
where $\vb{w}\in V$ is a vector (guaranteed to exist, since $(V,q)$ is regular) satisfying $q(\vb{w},\vb{v}_1)\neq0$.

We say that a quadratic space $(V,q)$ is an {\em orthogonal direct sum} and write $(V,q)=(U,q_u)\oplus(W,q_w)$ (or sometimes just $V = U\oplus W$ for brevity) if the vector space $V$ is a direct sum of vector spaces $U$ and $W$ and if $q(\vb{u},\vb{w})=0\,\forall\,\vb{u}\in U,\vb{w}\in W$. Every regular $(V,q)$ admits a \emph{Witt decomposition} into a direct sum $(V,q)=(V_h,q_h)\oplus(V_a,q_a)$, where $V_h$ is itself a direct sum of hyperbolic planes and $V_a$ is anisotropic. The \emph{Witt index} $w=\frac{1}{2}\dim V_h \in \mathbb{Z}$ is independent of the choice of decomposition and equals the dimension of 
any maximal totally isotropic subspace.

Given a Witt decomposition, it is easy to find a maximal
totally isotropic subspace:  simply pick one vector from each hyperbolic plane in $V_h$.
To effect a Witt decomposition, start from the Gram matrix whose diagonal entries are $c_i$ (ordered arbitrarily). A first hyperbolic plane $H_1$ can be found using any isotropic vector, such as $
\vb{v}_1=(1,1,\dots,1)$, 
along with a second one $\vb{v}_2$ determined using \cref{equation:v2}. By changing basis to the set containing $\vb{v}_1$, $\vb{v}_2$ and $n-2$ linearly independent vectors $\vb{v}$ satisfying $\vb{v}_1^\intercal\mathsf{Q}\vb{v}=\vb{v}_2^\intercal\mathsf{Q}\vb{v}=0$, the Gram matrix in the new basis will be the direct sum of $\begin{psmallmatrix}0&1\\1&0\end{psmallmatrix}$ (the Gram matrix of the hyperbolic plane) and an $(n-2)\times(n-2)$ matrix (not necessarily diagonal or traceless). Then iterate: search for an isotropic vector of the $(n-2)$-dimensional subspace, construct a hyperbolic plane, and decompose until $w$ hyperbolic planes have been found. Undoing the basis transformations yields an explicit Witt decomposition $V=H_1\oplus\dots\oplus H_w\oplus V_a$ of the original quadratic space.

{\em Generalized orthogonal transformations.}
Given just one maximal totally isotropic subspace, it is possible, using symmetry, to find all such spaces without repeating the procedure just described {\em ad nauseam}. Given quadratic spaces $(V_1,q_1)$ and $(V_2,q_2)$ over the same field $\Bbbk$, an isometry is an isomorphism of vector spaces $\sigma:V_1\to V_2$ such that $q_2(\sigma\vb{v})=q_1(\vb{v})\,\forall\,\vb{v}\in V_1$.
The isometries with $V_1 = V_2 = V$ form a group $O(V)$ under composition, which we call the \emph{generalized orthogonal group}. It acts transitively on the totally isotropic subspaces of any given dimension, so we can find all maximal totally isotropic subspaces from just  one, if we can somehow generate each element $\mathsf{O}$ of $O(V)$. 

In terms of our original Gram matrix $\mathsf{Q}$, an $\mathsf{O} \in O(V)$ can be represented as an $n\times n$ matrix (which we also denote $\mathsf{O}$) satisfying $\mathsf{O}^\intercal\mathsf{Q}\mathsf{O}=\mathsf{Q}$. This equation, being quadratic over $n(n-1)/2$ rational variables, is even harder to solve than \cref{equation:quaddiophantine}, so consider instead 
a \emph{generalized skew-symmetric matrix} $\mathsf{A}$ satisfying ${\mathsf{A}^\intercal\mathsf{Q}+ \mathsf{Q}\mathsf{A} = 0}$. This equation, being linear in $n(n-1)/2$ rational variables, is easily solved symbolically in \texttt{Mathematica}. Moreover, there is a 1-1 correspondence between $\mathsf{A}$ and the $\mathsf{O}$ which do not have 1 as an eigenvalue, given by a generalization of the usual Cayley transform~\cite{GantmakherF.R.FeliksRuvimovich2000Ttom}
\sansmath
\begin{align*}
	A=(I+O)(I-O)^{-1},&&O=-(I-A)(I+A)^{-1},
\end{align*}
\unsansmath
where $\mathsf{I}$ is the $n\times n$ identity matrix.

Thus we generate all generalized orthogonal matrices $\mathsf{O}$ not having 1 as an eigenvalue. To generate the remaining ones, note that, at least if $\mathsf{Q}$ is diagonal (as ours is), then $\mathsf{O'=DO}$ is generalized orthogonal if $\mathsf{O}$ is \footnote{Proof: $\mathsf{(DO)^\intercal QDO = O^\intercal D^\intercal QDO = O^\intercal QO = Q}$, where we have used the facts that $\mathsf{D}^\intercal=\mathsf{D}^{-1}=\mathsf{D}$ and that $\mathsf{D}$ and $\mathsf{Q}$ commute as they are both diagonal.}, where $\mathsf{D}$ is an $n\times n$ diagonal matrix with diagonal entries taken from $\{1,-1\}$.
Hence, we can strengthen the result of Ref.~\cite{27f6df08-b306-396c-b148-e0befbe5274c} to the following statement: given the parameterization of all generalized orthogonal matrices not having 1 as an eigenvalue, left-multiplying it by all $2^n$ possible $\mathsf{D}$ yields all generalized orthogonal matrices.

\section{An example}\label{section:ex}
Consider the six-dimensional quadratic form
$$x_1^2+3x_2^2+3x_3^2-2x_4^2-2x_5^2-3x_6^2$$
with Gram matrix $\mathsf{Q}=\mathrm{diag}(1,3,3,-2,-2,-3)$.
This could correspond physically to the abelian part of the local anomaly in an $SU(2) \times U(1)^m$ gauge theory whose left-moving fermions are two $SU(2)$ doublets and a triplet and whose right-moving fermions are a singlet and two triplets, with $x_i$ denoting the charges of a $U(1)$ factor.

Using the ancillary \texttt{Mathematica} notebook, one finds that the Witt index is 2, and that, with the following change of basis $\mathsf{B}$ leading to the new Gram matrix $\mathsf{B}^\intercal\mathsf{QB}$,
\begin{align*}
	\mathsf{B}=
	\begin{psmallmatrix*}[r]
		1 & \frac{1}{2} & 0 & 0 & 0 & 0 \\
		1 & -\frac{1}{2} & 0 & 0 & \frac{2}{3} & \frac{2}{3} \\
		0 & 0 & 1 & \frac{1}{6} & 0 & 0 \\
		1 & -\frac{1}{2} & 0 & 0 & 1 & 0 \\
		1 & -\frac{1}{2} & 0 & 0 & 0 & 1 \\
		0 & 0 & 1 & -\frac{1}{6} & 0 & 0 \\
	\end{psmallmatrix*},&&
	\mathsf{B^\intercal QB}=\begin{psmallmatrix*}[r]
		0 & 1 & 0 & 0 & 0 & 0 \\
		1 & 0 & 0 & 0 & 0 & 0 \\
		0 & 0 & 0 & 1 & 0 & 0 \\
		0 & 0 & 1 & 0 & 0 & 0 \\
		0 & 0 & 0 & 0 & -\frac{2}{3} & \frac{4}{3} \\
		0 & 0 & 0 & 0 & \frac{4}{3} & -\frac{2}{3} \\
	\end{psmallmatrix*},
\end{align*}
a maximal totally isotropic subspace of dimension 2 can be seen to be the span of the first and third columns of $\mathsf{B}$, {\em{i.e.}} $(1,1,0,1,1,0)^\intercal$ and $(0,0,1,0,0,1)^\intercal$ \footnote{The two-dimensional subspace spanned by the last two columns of $\mathsf{B}$ is anisotropic over the 5-adics and hence over $\mathbb{Q}$.}.
 
All generalized skew-symmetric matrices corresponding to $\mathsf{Q}$ are parameterized by the rational parameters $a_{1\, 2},\dots,a_{5\,6}$ as
$$\mathsf{A}=
\begin{psmallmatrix*}[r]
	0 & a_{1\, 2} & a_{1\, 3} & a_{1\, 4} & a_{1\, 5} & a_{1\, 6} \\
	-\frac{a_{1\, 2}}{3} & 0 & a_{2\, 3} & a_{2\, 4} & a_{2\, 5} & a_{2\, 6} \\
	-\frac{a_{1\, 3}}{3} & -a_{2\, 3} & 0 & a_{3\, 4} & a_{3\, 5} & a_{3\, 6} \\
	\frac{a_{1\, 4}}{2} & \frac{3 a_{2\, 4}}{2} & \frac{3 a_{3\, 4}}{2} & 0 & a_{4\, 5} & a_{4\, 6} \\
	\frac{a_{1\, 5}}{2} & \frac{3 a_{2\, 5}}{2} & \frac{3 a_{3\, 5}}{2} & -a_{4\, 5} & 0 & a_{5\, 6} \\
	\frac{a_{1\, 6}}{3} & a_{2\, 6} & a_{3\, 6} & -\frac{2 a_{4\, 6}}{3} & -\frac{2a_{5\, 6}}{3}  & 0 \\
\end{psmallmatrix*}.
$$
From here all generalized orthogonal matrices $\mathsf{O}$ and thus a parameterization of all solutions can be found by left-multiplying any vector living in the above maximal totally isotropic subspace, which has the form $(\alpha,\alpha,\beta,\alpha,\alpha,\beta)^\intercal$ for some $\alpha,\beta\in\mathbb{Q}$, by $\mathsf{O}$, though it is too unwieldy to reproduce here. Just as an example, choosing $\mathsf{D}=\mathsf{I}$ and setting $a_{1\,2},\dots,a_{5\,6}=1$ gives another maximal totally isotropic subspace spanned by $(7,9,10,10,10,8)^\intercal$ and $(147,49,0,120,0,-2)^\intercal$ which only intersects the previous subspace at the origin.

As we are able to parameterize all solutions by means of our algorithm, it is perhaps moot to compare it with a brute-force trial-and-error scan, which can only ever find a finite number. Each of the steps in our algorithm, 
namely finding the Witt index, obtaining a basis for a Witt decomposition, parameterizing generalized orthogonal matrices via generalized skew-symmetric ones, and generating all maximal totally isotropic subspaces from one such subspace, leads to a dramatic reduction in the scaling of the computational problem in $n,m$ and the allowed ranges of the charges. As an illustration, the evaluation of the {\tt Mathematica} notebook on the above example took less than a second on a laptop. By contrast, even after restricting the charges $x_1,\dots,x_6$ to values in $\{-1,0,1\}$, a scan using a basic {\tt Mathematica} script took 4.14s (resp. 49.86s) to carry out the $3^{12}$ (resp. $3^{18}$) calculations required to establish that there are 5184 (resp. 0) solutions with $m=2$ (resp. $m=3$) gauge fields.  

\acknowledgments
We thank Michael Brownlie for collaboration at an early stage of the project. This work was partially supported by STFC consolidated grants ST/T000694/1 and ST/X000664/1 and a Trinity-Henry Barlow Scholarship.

  \bibliography{2d_anomalies_submission_references_jhep}    
  \end{document}